\documentclass[review]{elsarticle}

\usepackage{hyperref,color}

\usepackage{array}
\usepackage{amssymb}
\usepackage{extarrows}
\usepackage{graphicx}
\usepackage[normalem]{ulem}
\usepackage{multirow}
\usepackage[table]{xcolor}
\usepackage{algorithm}
\usepackage{algorithmic}
\usepackage{subfig}

\usepackage{xcolor}
\usepackage{tabularx}

\usepackage{url}
\usepackage{endnotes}

\usepackage{float}

\usepackage[section]{placeins}

\makeatletter
\g@addto@macro{\UrlBreaks}{\UrlOrds}
\makeatother

\biboptions{authoryear}

\begin{document}

\begin{frontmatter}

\title{Computing Conceptual Distances between Breast Cancer Screening Guidelines:\\
An Implementation of a Near-Peer Epistemic Model of Medical Disagreement.
\tnoteref{}}
\tnotetext[mytitlenote]{$^*$ Corresponding authors:  
\{wzadrozn,hhematia,sgopala4\}@uncc.edu;Luciana.Garbayo@ucf.edu }

\author{Hossein Hematialam$^{*c}$,  Luciana Garbayo$^{*f}$, Seethalakshmi Gopalakrishnan$^{*c}$, Wlodek Zadrozny$^{*cd}$ \fnref{myfootnote}}
\address{$^c$ College of Computing, UNC Charlotte, Charlotte NC\\
$^d$ School of Data Science, UNC Charlotte, Charlotte NC\\
$^f$ Departments of Philosophy  \& Medical Education, U. of Central Florida, Orlando, FL
}





\begin{abstract}
Using natural language processing tools, we investigate the differences of recommendations in medical guidelines for the same decision problem -- breast cancer screening.
We show that these differences arise from knowledge brought to the problem by different medical societies, as reflected in the conceptual vocabularies used by the different groups of authors.
The computational models we build and analyze agree with the near-peer epistemic model  of expert disagreement proposed by Garbayo. 
Even though the article is a case study focused on one set of guidelines, the proposed methodology is broadly applicable.

In addition to proposing a novel graph-based similarity model for comparing collections of documents, we perform an extensive analysis of the model performance. In a series of a few dozen experiments, in three broad categories, we show, at a very high statistical significance level of 3-4 standard deviations for our best models, that the high similarity between expert annotated model and our concept based, automatically created, computational models is not accidental. Our best model achieves roughly 70\% similarity. We also describe possible extensions of this work.

\end{abstract}

\end{frontmatter}


\section{Introduction and Motivation}\label{sec:intro}

\textbf{Research Objective:}
In this article we investigate the differences in medical guidelines in response to the same decision problem: whether to recommend a breast cancer screening for patients with same conditions.  Our research objective is to create a computational model accurately representing medical guidelines disagreements; a model which is simple and general enough to be potentially applicable in other situations. We evaluate our approach using a case study, where we are asking whether differences in medical recommendations come from differences in knowledge\footnote{by knowledge we mean both the domain knowledge and the associate epistemic practices} brought to the problem by different medical societies.

This article should also be viewed as a case study in computational implementation of the near-peer epistemic model of expert disagreement proposed in several of our earlier work (\cite{garbayo2014}, \cite{Garbayo2018}, \cite{garbayoPrague}, \cite{garbayo2019dependence}).
The near-peer model can be viewed as a refinement of the standard epistemic peer model (e.g. \cite{lackey2014rel}).

\begin{figure}[ht]
\includegraphics[width=1.0\textwidth]{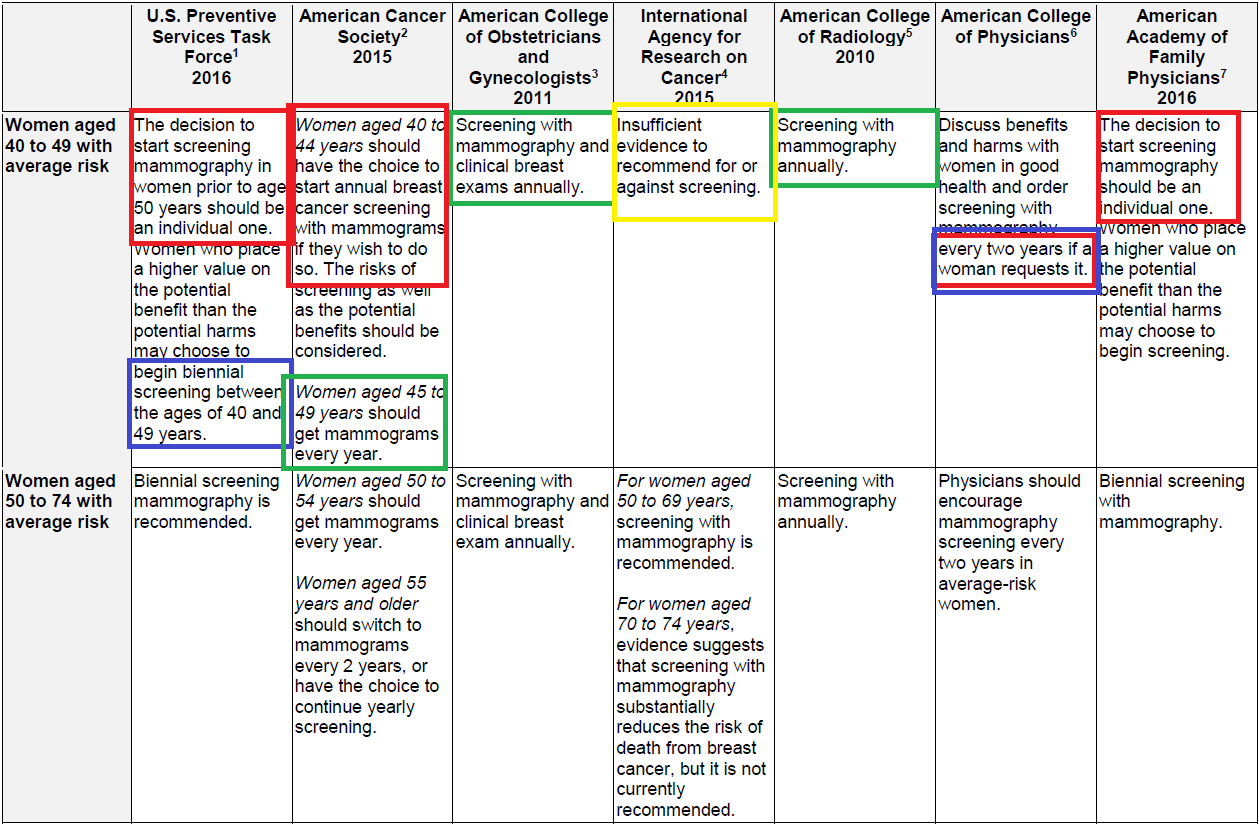}
\caption{Note the contradictory recommendations in green and blue boxes. In \cite{ZadroznyIWCS2017}, we annotated the table from CDC comparing recommendations of different medical societies \cite{CDC_2017} to explicitly show the types of disagreements in different guidelines. We also showed there that some of these contradictions can be computed automatically. Only part of the table is reproduced here.}
\label{fig:cdc}
\end{figure}

More specifically, we use natural language processing tools to build computational representations of a set of seven breast cancer screening guidelines. These computational representations are created from \textit{full texts} of the guideline documents. Since we are using vectors to represent the documents, we can ask whether the distances between vectors (e.g. the cosine distance) are semantically significant; and in particular,  whether the degrees of \textit{conceptual disagreement} between the guidelines  correspond to the differences in semantic  distances automatically computed from full documents. We measure the conceptual disagreement using a CDC \textit{summary } of the full documents \cite{CDC_2017}, which focuses precisely on the differences in screening recommendations. An annotated summary of the seven documents from \cite{CDC_2017} is reproduced in Fig.\ref{fig:cdc}, with colors added for areas of agreement and  disagreement. In addition, we plot these disagreements in a diagram (see Fig.\ref{fig:lh}).

In other words, we ask about whether concepts, that is single words or two consecutive words (bigrams), which are present in full documents determine the  \textit{relations} seen in the documents' summary. That is, whether concepts used by experts affect their recommendations. (Note that the concepts that appear in summaries, such as mammography, are common in all documents, so the difference in recommendations must come from other knowledge).

Technically speaking, we compute semantic distances between several pairs of breast cancer screening guidelines, using different automated methods, and then we compare the resulting measurements to an expert opinion in Fig. \ref{fig:lh}. As we explain later in the article these comparisons suggest the potential for computational tools to be used to estimate epistemic distances between the guidelines.\\
  
\noindent
\textbf{Motivation:}   
Guidelines are complex products for medical decision making. 
Despite some convergence in the interpretation of medical evidence,
different medical specialties might produce dissimilar guidelines for the same medical problem. Nevertheless, physicians who have similar training tend to generate similar guidelines based on their sharing of methods, objects and overall academic background typical of the specialty \cite{Garbayo2018}.
These specialties cover a specific region of theoretical and practical medical knowledge that may overlap with others. For example, oncology or radiology overlaps with general practice. Per our hypothesis, this variable similarity of backgrounds and specialization is likely to be reflected in the types recommendations for either screening, treatment or prevention of specific conditions, such as breast cancer. 
 
Since medical guidelines are developed by different medical associations 

which count on experts with different specialties and sub-specialties, there is a high possibility that there may be disagreement in the guidelines. And indeed, as noted by \cite{CDC_2017} and discussed in \cite{zadrozny2018sheaf,ZadroznyIWCS2017}, breast cancer screening guidelines contradict each other. Besides breast cancer screening disagreements, which we model in this article, we have witnessed controversies over PSA screening, hypertension and other treatment and prevention guidelines. \\

\noindent
\textbf{The near-peer hypothesis:}  
Conceptually, this article presents a computational study in epistemic modeling of medical guidelines disagreement, as a model of \textit{near-peer }disagreement. By that we mean the following: traditionally, 
domains experts, e.g. the authors of medical guidelines, have been viewed as \textit{epistemic peers}, i.e. groups of professionals accessing the same medical knowledge in the process of creating the guidelines; and in the case of disagreements, they can evaluate opposing views, and have identical levels of competence (\cite{garbayo2019dependence}). As exemplified in \cite{christensenepistemology}, this has been the dominant paradigm in analyzing disagreements among experts (peers).

However, a more realistic and fine grained model is possible, namely, where we see these groups as having partly overlapping knowledge, and therefore can be named \textit{‘near-peers.’} While, intuitively, such a model sounds reasonable, it raises the question, how should such near-peer models be created and analyzed. Clearly we do not know exactly what kind of knowledge the individuals involved in  creating the guidelines bring to the table, and even with access to the discussions underlying the creation process,  doing a manual conceptual  analysis would be slow and tedious. Yet, given the progress achieved in building computational models of text documents \cite{zhou2020progress}, we can hypothesize that such computational models might be of some use. 
Hence this hypothesis:  

\textbf{Hypothesis: } The epistemic near-peer disagreement between medical experts and their societies can be measured using natural language processing techniques to measure the conceptual distance between the produced guidelines. \\

\noindent
\textbf{Contributions: }The main contribution of this article is in showing that automated, and relatively straightforward, methods of text analysis can compute conceptual differences between documents addressing the same topic (breast cancer screening); and these automated judgments have a moderate to high correlation with an expert judgment. Namely, we compute the similarity and the dissimilarity between the breast cancer guidelines provided by different medical societies, using a few standard methods of representing text and computing such metrics. We then correlate it with previously done conceptual analysis of the main recommendations of these guidelines. Thereby, we show the viability of the near-peer model.

Another contribution is the articulation of a very natural graph-clique based algorithm/method for comparing similarity of two \textit{collections} of documents. Given two sets of documents, each of the same cardinality, and a mapping between nodes, 
we compute the percent distortion between the shapes of the two cliques, and the chances that the mapping arose from a random process. 
\footnote{Given the naturalness of the method, it's likely that we are reinventing something, but we couldn't find anything similar in literature. We would appreciate pointers to related prior art}.

We also document all steps of the process and provide the data and the code\footnote{The Github link will be provided here before publication.}

to facilitate both extensions of this work and its replication. Even though NLP methods have progressed enormously in the last decade \cite{zhou2020progress}, they are far from perfect. In our experiments, we use some of the simplest semantic types words and simple collocations represented as vectors in high dimensional spaces. However, this simplicity is helpful, as we can run several experiments, and compare the effects of using different representations and metrics. This gives us confidence that the correlations we are discovering tell us something interesting about  guideline documents. 
 
Although the article merely establishes this correspondence in one case, nevertheless it might be a good starting point for analysis of other medical guidelines, and perhaps other areas of expert disagreement.  In addition, fast progress in automated document analysis using text mining and deep learning techniques can perhaps make such analyses more accurate and deeper.\\

\noindent
\textbf{Organization of the article: } 
In Section \ref{sec:priorart}, we provide both, an overview of applications of natural language processing to texts of medical guidelines, and introduce the near-peer model of epistemic disagreement. Section \ref{sec:lufig} explains our data sources: a CDC summary table of breast cancer screening guidelines and the corresponding full text documents. There, we also discuss the steps in the conceptual analysis of the table, resulting in a graph of conceptual distances between the columns of the table (i.e. summaries of the full documents).
We then proceed to the analysis of full documents using a two standard vectorization procedures in Section \ref{sec:distauto}. After observing roughly 70\% correlation between the distances in the summaries and the distances in the full documents, we prove in Section \ref{sec:results} that this correlation is not accidental.
We conclude in Sections \ref{sec:results} and \ref{sec:concls} that this case study shows that NLP methods are capable of approximate conceptual analysis, in agreement with the near-peer model. This opens the possibility of deepening such research using more sophisticated tools such as relationship extraction and automated formal analysis.

\section{Discussion of prior art}\label{sec:priorart}

We are not aware of any work directly addressing the issues we are tackling in this article; namely, the automated conceptual analysis of medical screening  recommendations, and connecting such analysis to broader problems of epistemic peers or near-peers and their disagreements. 
However, there is a body of knowledge addressing similar issues individually, which we summarize in this section.

\subsection{Text analysis of medical guidelines}\label{sec:TAmedG}

An overview article \cite{peek2015thirty},  from a few years ago, states that different types of analysis of medical guidelines are both a central theme in applications of artificial intelligence to medicine, and a domain of research with many challenges. The latter includes building formal, computational representations of guidelines and a wider application of natural language processing. From this perspective, our work is relevant to these central but general themes.

To switch to more recent and more technical work, 
\cite{bowles2019framework} focuses on finding and resolving conflicting recommendations using a formal model and automated proof systems -- it relies on a manual translation into a formal language, Labelled Event Structure. This is a very interesting work, somewhat in the spirit of our own attempts to do it, \cite{ZadroznyIWCS2017}, using a combination of NLP and information retrieval tools. 
In another article dealing with contradictory recommendations, 
\cite{tsopra2018using}  focus on the semi-automatic detection of inconsistencies in guidelines and apply their tools to antibiotherapy in primary care. 
In an application of Natural Language Processing, 
\cite{lee2020natural}
 show that one can accurately measure
adherence to best practice guidelines in a context of palliative care.

More broadly, modern NLP methods have been applied to 
clinical decision support, e.g. 
\cite{seneviratne2019enabling}, with ontologies and semantic web for concept representation; and to automatic extraction of adverse drug events and drug related entities, e.g. \cite{ju2020ensemble} using a neural networks model. For document processing, we have e.g.  \cite{benedetti2019computing} proposing a knowledge-based technique for inter-document similarity computation, and \cite{rospocher2019boosting} successfully applying conceptual representations to document retrieval.

All of these show show that the state-of-the-art systems are capable both of performing statistical analysis of large sets of documents, and a semantic analysis fitting the need of a particular application. 

\noindent
\textbf{Extending the limits of current practice:}
This work extends the state-of-the-art computational analysis of medical guidelines. Namely, instead of semi-automated conceptual analysis, we prove the feasibility of automated conceptual analysis. That is, in our study, we use a representation derived from a (relatively shallow) neural network (BioASQ embeddings \cite{BioASQ2015a}), and knowledge-based annotations derived from MetaMap \footnote{\url{https://metamap.nlm.nih.gov/}}. Our results, in Section \ref{sec:results}, show that both are useful as representations of our set of guidelines, and show similar performance in modeling conceptual similarities. From the point of view of methodology of analyzing medical guidelines, this article contains the first computational implementation of  the near-peer model.

\subsection{Conceptual analysis of disagreement in medical guidelines}\label{sec:CAmedG}

The significance of medical guidelines disagreement, such as in the case of breast cancer screening disagreement, is expressed in the projected preventable harms of unwarranted clinical variation  in health care, \cite{sutherland2019unwarranted}.
The cited article proposes a theoretical framework for the examination of unwarranted clinical variation, and suggests  we can identify such unwarranted clinical variation in three dimensions: “assessing variation across geographical areas or across providers;" through “criteria for assessment, measuring absolute variation against a standard, or relative variation within a comparator group"; and ”as object of analysis, using process structure/resource, or outcome metrics."

It is important to notice that such unwarranted clinical variation is not about precision medicine and customization of care; rather, it represents a disagreement that includes knowledge management and translation issues, as well as a variation of epistemic practices. For instance, \cite{solomon2015making}, provides a compelling social epistemology study of consensus conferences and their epistemic pitfalls. 

In this article we focus more narrowly on the epistemic analysis of medical guidelines disagreement. As argued in our earlier work (\cite{garbayoPrague}),
the expectation of a epistemic agreement and consensus across different medical societies of specialties rests in a historic habit of idealization. That is, it lies
in viewing domain experts as having identical knowledge and identical reasoning capabilities, and whose reasoning processes can -- in principle -- be expressed formally and mechanically verified. 

Departing from a broader interpretation of epistemic peers
\cite{lackey2014rel},
to allow for variation, 
\cite{garbayo2019diagnosis} and \cite{garbayoPrague} 
explore the \textit{de}-idealization of medical consensus and disagreement, and propose a category of near-peers, to express more accurately the mismatches in knowledge domain and variability among multiple guidelines developers as epistemic agents. In the broader interpretation, they are all epistemic peers, but, in the study of variations, such micro-analyses provide us conceptual distances between their perspectives in a measurable way.

\noindent
\subsection{The gap between theory and engineering analysis of contradictory guidelines}\label{sec:gap}

Based on the above discussion, there is a clear gap between theoretical analyses of the creation and content of medical guidelines and the computational analysis of the same content, , e.g. from the epistemic point of view. 
In particular, as far as we know, there has been no computational investigation of contradictory guidelines incorporating the epistemic point of view of either expert peers nor near-peers. We demonstrate in the subsequent sections the feasibility of such computational models.

\section{The approach: data sources of guidelines and conceptual analysis} \label{sec:lufig}

As mentioned earlier, we are comparing the result of a manual conceptual analysis of several breast cancer screening guidelines with an automated analysis. In this section, we describe the process of this manual analysis of the summary document. In particular we produce numerical representations of the differences in the guidelines (per \cite{CDC_2017}), which later in Section \ref{sec:distauto} will be compared with the results of an automated analysis of full guideline documents.\\

\noindent
\textbf{The guidelines documents: } In this article, we are using both the CDC summary (\cite{CDC_2017}, reproduced and annotated in Fig.\ref{fig:cdc}), and the full text of the guidelines used by the CDC. The detailed information about these guidelines is shown in Table \ref{tab:guideref}. The focus of this section is on Fig. \ref{fig:cdc}. 

\begin{table}[ht]
\arrayrulecolor{black}
\resizebox{\textwidth}{!}{%
\begin{tabular}{|p{2.5cm}|p{3.5cm}|p{6cm}|p{3cm}|l}
\cline{1-4}
 \textbf{Guideline Abbreviation}& \textbf{Full Name of the Organization}& \textbf{URL Reference} & \textbf{Document Citation} &  \\ \cline{1-4}
 \textbf{ACOG}& The American College of Obstetrics and Gynecology & \url{http://msrads.web.unc.edu/files/2019/05/ACOGBreastCAScreening2014.pdf} & \cite{ACOGPracticeBulletin}  &  \\ \cline{1-4}
 
 \textbf{AAFP}& American Academy of Family Physicians & \url{https://www.aafp.org/dam/AAFP/documents/patient_care/clinical_recommendations/cps-recommendations.pdf} & \cite{action2017summary} &  \\ \cline{1-4}
 
 \textbf{ACP}& American College of Physicians & \url{https://annals.org/aim/fullarticle/2294149/screening-cancer-advice-high-value-care-from-american-college-physicians} & \cite{wilt2015screening} &  \\ \cline{1-4}
 
 \textbf{ACR}& American college of Radiology & \url{https://www.sciencedirect.com/science/article/pii/S1546144009004803} & \cite{lee2010breast} &  \\ \cline{1-4}
 \textbf{ACS}& American Cancer Soceity & \url{https://www.ncbi.nlm.nih.gov/pmc/articles/PMC4831582/}  & \cite{oeffinger2015breast} &  \\ \cline{1-4}
 \textbf{IARC}& International Agency for Research on Cancer & \url{https://www.nejm.org/doi/full/10.1056/NEJMc1508733}  & \cite{jorgensen2015breast} &  \\ \cline{1-4}
 \textbf{USPSTF}& United States Preventive services Task Force  & \url{https://annals.org/aim/fullarticle/2480757/screening-breast-cancer-u-s-preventive-services-task-force-recommendation}  & \cite{siu2016screening}  &  \\ \cline{1-4}
\end{tabular}%
}
\caption{Guidelines with references}
\label{tab:guideref}
\end{table}

As shown in Fig.\ref{fig:cdc}, reproduced from our earlier work 
\cite{hematial2016s} and \cite{zadrozny2018sheaf},  there are several clear  disagreements in the recommendations. \\

\begin{figure}[ht]
\includegraphics[width=1.1\textwidth]{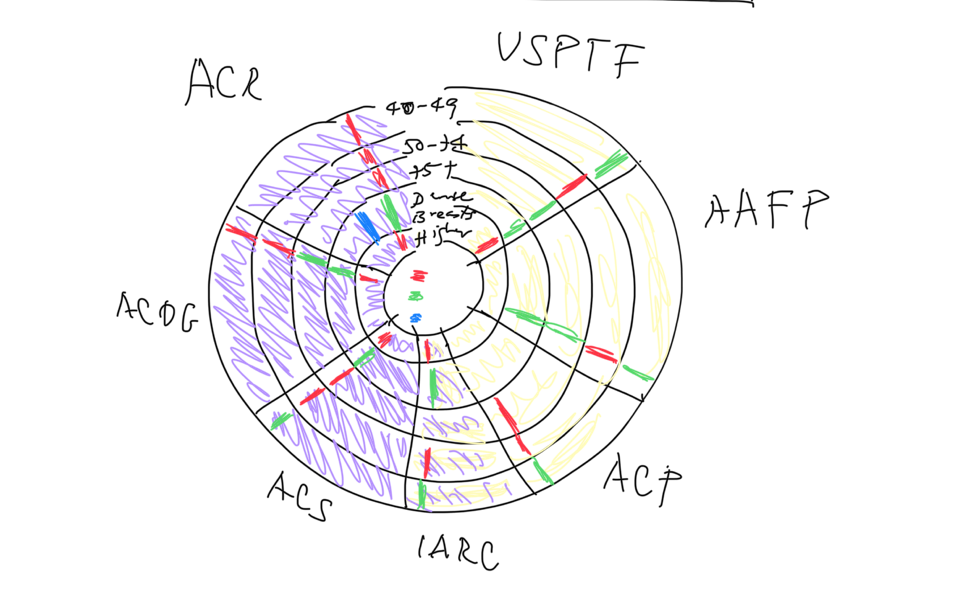}
\caption{Similarities between guidelines using expert annotations. The yellow coloring shows patient making decisions, the purple coloring shows explicit screening recommendations. The concentric circles refer show different age groups. Red marks -- physician decides, green marks -- patient decides.}
\label{fig:lh}
\end{figure}

\begin{table}[ht]
 \arrayrulecolor{black}
\resizebox{\textwidth}{!}{%
\begin{tabular}{|l|c|c|c|c|c|}
\hline
\textbf{Guideline} & \textbf{40-49} & \textbf{50-74} & \textbf{75+} & \textbf{Dense Breast} & \textbf{Higher than average risk} \\ \hline
AAFP            & b & r & b & b & N \\ \hline
\textbf{ACOG}   & r & r & b & b & r \\ \hline
\textbf{ACP}    & b & r & r & N & N \\ \hline
\textbf{ACR}    & r & r & r & b & r \\ \hline
\textbf{ACS}    & b & r & r & b & b \\ \hline
\textbf{IARC}   & b & r & N & b & r \\ \hline
\textbf{USPSTF} & b & r & b & b & r \\ \hline
\end{tabular}%
}
\caption{The table shows recommendations as follows: 
N --- no recommendation; b --- both, patient and doctor, shared decision; r --- recommending mammography. }
\label{tab:LuFeat}
\end{table}

\noindent
\textbf{Conceptual Analysis: } 
Figure \ref{fig:lh} \ shows a manually generated 

graph showing the differences between the guidelines, also presented in \cite{garbayo2019diagnosis}. There are two sides to the circle. The yellow side indicates the scenario where patients will likely decide when breast cancer screening should be done, and the purple color side specifies the situation where breast cancer guideline providers most likely will demand screening interventions. Black color indicates the different societies boundaries. The red color marks indicate the physician decides. Green color marks indicate patients' decisions. Since the consideration of ultrasound (blue) appears only in radiology guidelines, we decided to abstract it out. 

The observed differences seem to support a near-peer model:  we see partially overlapping agreements. 
If we look into the above diagram (Fig. \ref{fig:lh}) we can infer that ACS is the nearest guideline to IARC on the purple side and USPTF is the nearest one on the yellow side. USPSTF is the nearest guideline to the AAFP. ACR is the farthest node to USPSTF. ACR, ACOG are close to each other. Notice that all we did was to observe the differences in recommendations. We are not trying to judge if some of these differences are more important than the others. Obviously, this categorization process is informed by medical knowledge, but the lists of differences are clear from reading the document reproduced in Fig.\ref{fig:cdc}.
So the question is whether we can reproduce these similarities and differences using a fully automated process, and without access to the summary document in Fig.\ref{fig:cdc}.

\FloatBarrier
{Table \ref{tab:LuFeat} represent the content of this analysis as a collection of features. Table \ref{tab:LuDist} shows the distances between the guidelines derived from Tables \ref{tab:LuFeat} and \ref{tab:featdiff} using the Jaccard distance (the percentage of different elements in two sets):
\[d_{j}(A,B)=1 - \dfrac{\mid A  \cap B \mid}{\mid A  \cup B \mid}\]
}

\FloatBarrier
\bigskip
\begin{table}[!h]
\resizebox{\textwidth}{!}{%
\begin{tabular}{|c|c|c|c|c|c|c|c|}
\hline
\ & \textbf{AAFP} & \textbf{ACOG} & \textbf{ACP} & \textbf{ACR} & \textbf{ACS} & \textbf{IARC} & \textbf{USPSTF} \\ \hline
\textbf{AAFP}   & 0 & 2 & 3 & 3 & 2 & 2 & 1 \\ \hline
\textbf{ACOG}   & 2 & 0 & 4 & 1 & 2 & 2 & 1 \\ \hline
\textbf{ACP}    & 3 & 4 & 0 & 3 & 2 & 3 & 3 \\ \hline
\textbf{ACR}    & 3 & 1 & 3 & 0 & 1 & 2 & 2 \\ \hline
\textbf{ACS}    & 2 & 2 & 2 & 1 & 0 & 1 & 1 \\ \hline
\textbf{IARC}   & 2 & 2 & 3 & 2 & 1 & 0 & 1 \\ \hline
\textbf{USPSTF} & 1 & 1 & 3 & 2 & 1 & 1 & 0 \\ \hline
\end{tabular}%
}
\caption{This table shows the number of different feature values for pair of guidelines, based on Table \ref{tab:LuFeat}.}
\label{tab:featdiff}
\end{table}
\noindent
\begin{table}[h]
\resizebox{\textwidth}{!}{%
\begin{tabular}{|c|c|c|c|c|c|c|c|}
\hline
 & \textbf{AAFP} & \textbf{ACOG} & \textbf{ACP} & \textbf{ACR} & \textbf{ACS} & \textbf{IARC} & \textbf{USPSTF} \\ \hline
\textbf{AAFP}   & 0      & 0.0238 & 0.0357 & 0.0357 & 0.0238 & 0.0238 & 0.0119 \\ \hline
\textbf{ACOG}   & 0.0238 & 0      & 0.0476 & 0.0119 & 0.0238 & 0.0238 & 0.0119 \\ \hline
\textbf{ACP}    & 0.0357 & 0.0476 & 0      & 0.0357 & 0.0238 & 0.0357 & 0.0357 \\ \hline
\textbf{ACR}    & 0.0357 & 0.0119 & 0.0357 & 0      & 0.0119 & 0.0238 & 0.0238 \\ \hline
\textbf{ACS}    & 0.0238 & 0.0238 & 0.0238 & 0.0119 & 0      & 0.0119 & 0.0119 \\ \hline
\textbf{IARC}   & 0.0238 & 0.0238 & 0.0357 & 0.0238 & 0.0119 & 0      & 0.0119 \\ \hline
\textbf{USPSTF} & 0.0119 & 0.0119 & 0.0357 & 0.0238 & 0.0119 & 0.0119 & 0      \\ \hline
\end{tabular}%
}
\caption{Distances between the summarized guidelines computed using Jaccard distance from Tables \ref{tab:featdiff} and \ref{tab:LuFeat}}
\label{tab:LuDist}
\end{table}

\begin{figure}[]
\centering

\subfloat[Jaccard distances on annotated recommendations, as per Table \ref{tab:LuFeat}. \label{fig:l1a}]{\includegraphics[width=0.44\textwidth]{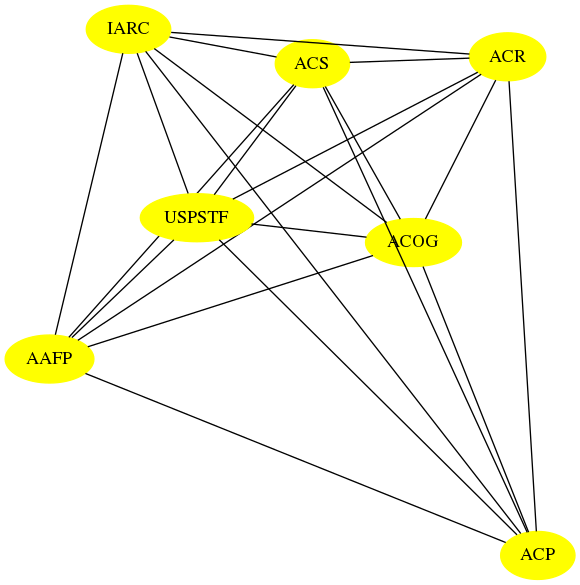}}\hfill

\subfloat[Number of differing features.\label{fig:l1b}] {\includegraphics[width=0.55\textwidth]{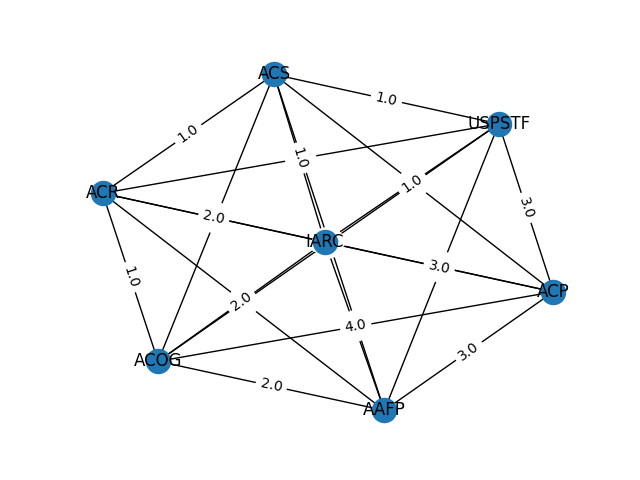}}\hfill

\caption{A pictorial representation of the distances between  recommendations, and numbers of differing features, as per Tables \ref{tab:featdiff} and \ref{tab:LuDist}. Can we replicate this geometric structure using automated tools? See Section \ref{sec:distauto} for an answer. }\label{fig:l1}
\end{figure}

\FloatBarrier
Tables \ref{tab:LuFeat}, \ref{tab:featdiff}  and \ref{tab:LuDist} represent the process of converting the information in Fig. \ref{fig:lh} into a set of distances. 
These distances are depicted graphically in 
Fig. \ref{fig:l1}, where we depict both Jaccard distances between the annotated guidelines, and the number of differing features as per Table \ref{tab:featdiff}.
\FloatBarrier

In Section \ref{sec:distauto}, we investigate how close automated tools can replicate this analysis, using  document distances typically used in information retrieval and natural language processing.

 \FloatBarrier
 
\section{Automated analysis of conceptual distances between document guidelines} \label{sec:distauto}

In\ the last 10 years, we have witnessed a new era in automated semantic analysis of textual documents \cite{zhou2020progress}. While no system can claim to ‘really’ understand natural language, in several domains such as data extraction, classification and question answering, automated systems dramatically improved their performance, and in some cases perform better than humans, due to the unmatched pattern recognition and memorization capabilities of deep neural networks (see e.g. \cite{smith2020contextual} for an overview).  

Some the simplest, easiest to use and effective of the new methods are different types of word and concept embeddings   (\cite{mikolov2013distributed}, \cite{pennington2014glove}, \cite{Shalaby2018irj}, \cite{kalyan2020secnlp}). Embeddings represent words and concepts as dense vectors (i.e. all values are non-zero), and are a preferred tool to make similarity judgments on the level of words, phrases, sentences and whole documents. 

Word embeddings have been widely used to compare documents, and in particular to compute their degree of similarity \cite{nguyen2019learning,tien2019sentence}.
Other methods proposed to compute documents similarity are based on using background knowledge \cite{benedetti2019computing}. This works uses both methods, human knowledge encoded the analysis of the CDC table, and embeddings.

However, before we create our document embeddings Section \ref{sec:docembed} and show the similarity between the representations of full documents and their summaries in Section \ref{sec:bestmodel}, a few words about data preparation are in order.

\subsection{Data preparation for all experiments}\label{sec:dataprep}

\noindent 
From the breast cancer screening guidelines listed in the CDC summary document \cite{CDC_2017}, USPSTF, ACS, ACP, and ACR guidelines are available in the HTML format, from which we extracted the texts of these guidelines. We used Adobe Acrobat Reader to obtain the text\  from the pdf format of AAFP, ACOG, and IARC guidelines. Since the AAFP documents also included preventive service recommendations for other diseases (such as other types of cancers), we added a preprocess step to remove those recommendations, and leaving the parts matching ‘breast cancer’.

As mentioned earlier, the manually annotated distances were obtained from the CDC table (referenced above in Fig. \ref{fig:cdc}), which \textit{summarized} all the breast cancer guidelines. However, the automated computation of conceptual distances was performed on the \textit{full} guideline documents. \\

\noindent
\textbf{Additional Experiments: }We also performed additional experiments with \textit{modified} views of the full guidelines documents, as enumerated below. This was driven by the fact that the levels of agreement may change if we compute the similarities/distances between selected sentences, which are explicitly related to the statements from the CDC table in Fig. \ref{fig:cdc}. For these additional experiments we split each guideline document into two different subsets: 
\begin{enumerate}
    \item \texttt{Related:} containing sentences that are related to CDC table, by having common concepts, as represented by UMLS concepts.
    This was done in multiple ways, giving us 6 possible experiments: 
    \begin{enumerate}
        \item All the sentences in the CDC guideline table were considered as a single document. If a sentence had a  number of mutual concepts with that document, that sentence was considered as related sentence.
        \item If a sentence had \textit{minimum} number of mutual concepts with at least one statement from CDC table, that sentence was considered as related sentence.
    \end{enumerate}
     Different minimum numbers of mutual concept(s) were examined in our experiment, that is the \textit{ minimum} was set at  1, 2, and 3.
    \item \texttt{Unrelated:} the other sentences.

\texttt{Unrelated} sentences were not used for these additional experiments. 
\end{enumerate}

\bigskip
\noindent

For full text guidelines (as per Table \ref{tab:guideref}), the result of the experiments are shown in Table \ref{tab:featdiff2}, are discussed in Sections \ref{sec:align} and \ref{sec:concls}.
 For full text guidelines minus \texttt{Unrelated } sentences, the result of this experiment in Tables \ref{tab:Err1} and \ref{tab:Err2}, discussed in Sections \ref{sec:addExperiments} and \ref{sec:concls}. \\

\noindent 

\textbf{Concept extraction: }For all experiments, we used MetaMap\footnote{\url{ 
 metamap.nlm.nih.gov/}}
 to extract UMLS concepts \footnote{UMLS Concept: \url{
 https://www.nlm.nih.gov/research/umls/index.html}} and semantic types
\footnote{\url{https://www.nlm.nih.gov/research/umls/META3_current_semantic_types.html}} 
  in sentences. We only considered concepts with informative, in our opinion, semantic types. This meant using concepts related to diagnosis and prevention, for example 'findings,' and not using ones related e.g. to  genomics. Our final list had the following: [[diap], [hlca], [dsyn], [neop], [qnco], [qlco], [tmco], [fndg], [geoa], [topp], [lbpr]].

\subsection{Vector representations and similarity measurements used in the experiments}\label{sec:docembed}

After data preparation, 
our approach consists of 
using a vector representation of each document (guideline), and  measuring  similarities (or, equivalently, distances) between each pair of the vectors 
representing the documents. We use two standard measures:  
cosine similarity and word mover's distance (WMD, WM distance) \cite{kusner2015word}\footnote{In addition to cosine and WMD, we have also experimented with other metrics, and other views data such as\  search score, search rank, obtaining results in agreement with the ones reported in the article. 
}; we use Gensim (\cite{vrehuuvrek2011gensim}) as a tool for our experiments.

We experimented with three language models of medical guidelines disagreement: "no concept," conceptualized and BioASQ (see Tables \ref{tab:featdiff2}, \ref{tab:Err1} and \ref{tab:Err2}).
The first two were trained using the PubMed articles as the training data.   
The third one used pre-trained BioASQ word embeddings 
created for the BioASQ competitions \cite{BioASQ2015a}.\footnote{\url{http://BioASQ.org/news/BioASQ-releases-continuous-
space-word-vectors-}\\\url{obtained-applying-word2vec-pubmed-abstracts}}

Our first model, trained on PubMed includes only words (no additional conceptual analysis with MeSH\footnote{\url{https://www.nlm.nih.gov/mesh/meshhome.html}}
was done). In the second,
 more complex model,  MeSH   terms are replaced with n-grams.
  For example, if \texttt{breast}  and \texttt{cancer} appeared next to each other in the text, they are replaced with \texttt{breast-neoplasms} and treated as a concept.\\

The details of our experiments, with computation steps and algorithms to get numerical values, are shown and discussed in Sections \ref{sec:align}, \ref{sec:graphAlg} and \ref{sec:addExperiments}. Because of the large number of experiments we performed it might be best to discuss our best model, before going into the gory details of the experiments.

\noindent 
\subsection{Our best model: Using BioASQ embeddings and word mover's  distance}\label{sec:bestmodel} 

\begin{figure}[htbp]
\centering
\subfloat[Distances between the seven guidelines, based on human annotations.\label{fig:LuGraph2}]{\includegraphics[width=0.48\textwidth]{media/Distance_Graph_LU.png}}\hfill
\subfloat[Our best distance results, based on WM distance and BioASQ embeddings with concepts(see text for explanations) .\label{fig:wmdGraph}]{\includegraphics[width=0.48\textwidth]{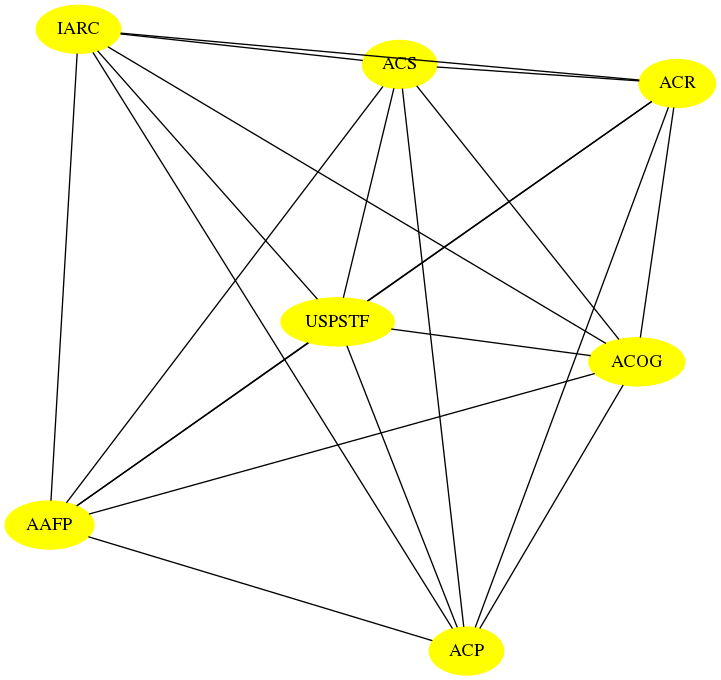}}
\caption{Visual comparison of a the similarity/distance graphs based on human analysis in panel (a), and computer generated from Table \ref{tab:bstdis} in panel  (b) suggests similar geometry. 
As we show in Section \ref{sec:results} this 69\% similarity is not accidental; the distortion is about 31\% (see Table \ref{tab:featdiff2}). } \label{fig:graphcompare}
\end{figure}

Table \ref{tab:bstdis} shows (unnormalized) WM distances between the seven guidelines using BioASQ embeddings. Fig.  \ref{fig:graphcompare} shows side by side the geometries of the two graphs: one generated based on human comparisons of the abstracted guidelines, and the second one based on the machine generated representations of the full guideline documents. The similarity is visible in a visual inspection, and will be quantified in the next Section to be about 70\%. There, we will also answer two questions:
\begin{itemize}
    \item[--] How do we measure the distortion between the two graphs?
    \item[--] Could this similarity of shapes be accidental?
\end{itemize}

To create Fig. \ref{fig:graphcompare}, for each metric, a diagram representing the distance between the nodes (guidelines) and a diagram with the labelled edges were drawn, using Python networkx library. \footnote{\url{https://networkx.github.io/}}  All the values were normalized to the same scale to allow visual comparison.

\begin{table}[ht]
\resizebox{\textwidth}{!}{%
\begin{tabular}{|c|c|c|c|c|c|c|c|}
\hline
\ & \textbf{AAFP} & \textbf{ACOG} & \textbf{ACP} & \textbf{ACR} & \textbf{ACS} & \textbf{IARC} & \textbf{USPSTF} \\ \hline
\textbf{AAFP}   & 0. & 1.83395352 & 1.90306464 & 1.99483722 & 1.86600794 & 2.15345843 & 1.6818018 \\ \hline
\textbf{ACOG}   & 1.83395352 & 0. & 1.64927636 & 1.29021522 & 1.33306188 & 1.77360488 & 1.28616845 \\ \hline
\textbf{ACP}    & 1.90306464 & 1.64927636 & 0. & 1.85617147 & 1.66757977 & 1.95600257 & 1.67437544 \\ \hline
\textbf{ACR}    & 1.99483722 & 1.29021522 & 1.85617147 & 0. & 1.410209 & 1.87369102 & 1.38540442 \\ \hline
\textbf{ACS}    & 1.86600794 & 1.33306188 & 1.66757977 & 1.410209 & 0. & 1.67692863 & 1.1636015 \\ \hline
\textbf{IARC}   & 2.15345843 & 1.77360488 & 1.95600257 & 1.87369102 & 1.67692863 & 0. & 1.75375893 \\ \hline
\textbf{USPSTF} & 1.6818018 & 1.28616845 & 1.67437544 & 1.38540442 & 1.1636015 & 1.75375893 & 0. \\ \hline
\end{tabular}%
}
\caption{\textit{This table shows the words mover distances between guidelines using BiaAsq embeddings. This is our best model. }}
\label{tab:bstdis}
\end{table}

\FloatBarrier
 
\section{Details of experiments and analysis of results}\label{sec:results}

With any of the simple semantic metrics discussed above, we obtained higher than random alignment between the expert  and the  machine’s  judgments  of conceptual similarity of the guidelines documents. Our program did not perform the same kind of analysis as the expert; namely, there was no relationship extraction, no focus on who makes the decision etc. (see Section \ref{sec:lufig} above). The automated analysis used latent factors, that is, the statistical distribution of concepts in the documents was used to establish degrees of similarity.

In this section we first discuss the statistical properties of the experiments to show our models capture statistically significant geometric correspondences. Then we provide the details of the steps we used to obtain the geometry and the statistics. In the last subsections we show results of additional experiments where the \texttt{Unrelated} sentences were removed from full guidelines (per Section \ref{sec:dataprep}). 

\begin{table}[ht]
\resizebox{\textwidth}{!}{%
\begin{tabular}{|c|c|c|c|}
\hline
\textbf{Model} & \textbf{Distortion} & \textbf{Distortion of permutations} & \textbf{STD} \\ \hline
&  distances measured as $ 1-sim $ &  & \\ \hline
\textbf{BioASQ\_WMD} & 0.313933661 & 0.381378177 & 0.009017982\\ \hline
\textbf{Conceptualized\_WMD} & 0.335044003 & 0.391185128 & 0.009293257\\ \hline
\textbf{NoConcept\_WMD} & 0.344571557 & 0.388227188 & 0.009099648\\ \hline
\textbf{BioASQ\_CosineSim} & 0.417871068 & 0.595697672 & 0.015729293\\ \hline
\textbf{Conceptualized\_CosineSim} & 0.534525231 & 0.613500756 & 0.016266786\\ \hline
\textbf{NoConcept\_CosineSim} & 0.51399564 & 0.590931627 & 0.015386531\\ \hline
\textbf{Search} & 0.543647172 & 0.619579945 & 0.017539478\\ \hline
 &  &  & \\ \hline
 & distance measured as $ 1/(sim -1) $ &  & \\ \hline
\textbf{BioASQ\_WMD} & 0.313933661 & 0.381378177 & 0.009017982\\ \hline
\textbf{Conceptualized\_WMD} & 0.335044003 & 0.391185128 & 0.009293257\\ \hline
\textbf{NoConcept\_WMD} & 0.344571557 & 0.388227188 & 0.009099648\\ \hline
\textbf{BioASQ\_CosineSim} & 0.393430546 & 0.571706079 & 0.0149424\\ \hline
\textbf{Conceptualized\_CosineSim} & 0.476975323 & 0.558498927 & 0.014585966\\ \hline
\textbf{NoConcept\_CosineSim} & 0.478890935 & 0.55465835 & 0.014345847\\ \hline
\textbf{Search} & 0.327758835 & 0.374535374 & 0.008806565\\ \hline
\end{tabular}%
}
\caption{{This table shows the 
the values obtained in multiple experiments. 
Column 2, \texttt{Distortion}, shows the distortions of graphs produced using corresponding models from from Column 1.  
Average distortion of per permutation is shown in Column 3. 
\texttt{STD} is standard deviation of the distortion per permutation.  Note that the distortion is somewhat depended on the how we measure distances; however, the the shapes of the distributions are very similar.} }
\label{tab:featdiff2}
\end{table}

\subsection{Automated judgements significantly align with the expert judgement}\label{sec:align}
Table \ref{tab:featdiff2} shows the results of the experiments with full text of the guidelines. 
Given seven documents, and the similarities/distances from the features established by the expert (Table \ref{tab:LuDist}), the average distortion value, computed in over ten thousand simulations, is 0.523$\%$  where we always assume distance of 0 between a document and its copy. (See the diagonal of Table \ref{tab:LuDist}). 

For our best model, BioASQ\_WMD, we have a 69\% similarity.
As we can see in Table \ref{tab:featdiff2} the distortion of this model is about 31\%, the average distortion of permutations (using the distances produced by BioASQ\_WMD) is 38\%, however the standard deviation of the distortions is less than 1\%.
So, the distance between the our model and the mean is about 7 standard deviations. Therefore, we conclude the correlation between the shapes of the two graphs is extremely unlikely to be coincidental. Hence the model represents a non-trivial similarity. 

Moreover, we performed the same kind of analysis using different models, i.e. different embeddings and different measures. And while the distances and distortions change, the chances of similarities arising by accident are always smaller than 1/1000 (four standard deviations from the mean of distortions). By this standard statistical criterion, no matter what measures of distance we use, the similarity between two graphs, one from human annotations and the other from automated concept modeling, is non-trivial and not accidental. 
We conclude that vector based representation are capable of detecting conceptual differences, i.e. the types and densities of concepts brought to the writing of medical recommendations (at least in our case study). 

\subsection{Graph-based method for comparing collections of documents}\label{sec:graphAlg}

We use a very natural, graph-clique based method for comparing similarity of two \textit{collections} of documents. Given two sets of documents, each of the same cardinality, and a mapping between nodes, 
we compute the percent distortion between the shapes of the two cliques, and the chances that the mapping arose from a random process. In our case the nodes of both graphs have the same names (the names of the medical guidelines), but the shapes of the graphs are different, one coming from human comparisons (Fig. \ref{fig:cdc}) and the other from machine produced similarities/distances. The details of the method, which was used to produce results of the previous subsection (\ref{sec:align}) are below, as a list of steps with references to three simple algorithms listed after the steps. 

Specific steps to establish the conclusion that automated judgements \textit{significantly} align with the expert judgments: 
\begin{enumerate}
    \item We work with the full text guideline documents, prepared as described in Section \ref{sec:dataprep}
    \item We start with building a vectorial representations for each text document, based on one of the word/concept embeddings described in Section \ref{sec:dataprep}.
    \item Using WM distance  (or cosine similarity) we compute the distances between each of the vectors from the previous step.
    \item We put the labels and distances into an adjacency matrix $\mathcal{A}_G$ (using Algorithm \ref{algo:AG})
    \item Using the procedure of Algorithm \ref{algo:distort} we compute the distance/distortion between the two labeled graphs, using the matrix obtained in the previous step, and the matrix in Tab. \ref{tab:LuDist}. For our best model it is 0.31.
    \item We ask the question: could this distortion be accidental? I.e. could another permutation of the graph nodes produce a similar result, that is, match to a large degree the geometry of the graph created from human annotations, Fig. \ref{fig:LuGraph2}?
    
    \item To answer this question, we compute the average distortion and the standard deviation, based on all possible permutation of nodes ( $5040=7!$ permutations). The pseudo-code for this computation is in Algorithm \ref{algo:graphstat}.
    \item Based on the fact that, per Table \ref{tab:featdiff2}, the difference between our results and average distortion is  seven (or more) standard deviations, we conclude the that the matching of the two geometries is not accidental and is highly significant. 
\end{enumerate}

\FloatBarrier
\begin{algorithm}
 \caption{\textbf{ \ Computing Graph of Distances Between Guideline Documents}.  The output of Algorithm \ref{algo:AG} is shown in Fig. \ref{fig:l1} }
 \label{algo:AG}
 \begin{algorithmic}[1]
 \renewcommand{\algorithmicrequire}{\textbf{Input:}}
 \REQUIRE
 \texttt{Guidelines}: a set of guideline documents in textual format .\\
\texttt{ Model}: a model to compute distances between two documents.\\
 \renewcommand{\algorithmicensure}{\textbf{Output:}}
 \ENSURE  $\mathcal{A}_G$ --- Adjacency matrix of distances between document guidelines.
  \FOR {each pair of documents in \texttt{Guidelines}}
  \STATE Compute the \texttt{distance} between the documents according to \texttt{Model}
  \STATE Put the \texttt{distance} in  $\mathcal{A}_G$ 
  \ENDFOR
  \RETURN $\mathcal{A}_G$ 
 \end{algorithmic} 
 \end{algorithm}
 
\begin{algorithm}
 \caption{\textbf{Distance or Percentage Distortion between Two Complete Graphs (cliques of the same size).}\\ Note. For example, the distance between the two graphs in Fig.\ref{fig:graphcompare} is 0.31, equivalent to 31\% distortion}
 \label{algo:distort}
 \begin{algorithmic}[1]
 \renewcommand{\algorithmicrequire}{\textbf{Input:}}
 \renewcommand{\algorithmicensure}{\textbf{Output:}}
 
 \REQUIRE Adjacency Matrices $\mathcal{A}_1$, $\mathcal{A}_2$  
 of equal dimensions\\
 \ENSURE  Graph distance/distortion \ $ \mathcal{D}(\mathcal{A}_1, \mathcal{A}_2)$, \ as a value between $0$ and $1$.
  \STATE Normalize the distances in $\mathcal{A}_1$ (by dividing each distance by the sum of distances in the graph) to produce a new adjacency matrix $\mathcal{AN}_1$
  \STATE Normalize the distances in $\mathcal{A}_2$  to produce a new adjacency matrix $\mathcal{AN}_2$
  \STATE Set the value of $graph\_distance$ to $0$.
  \FOR {each \texttt{edge} in $\mathcal{AN}_1$}
  \STATE Add the absolute value of the difference between the \texttt{edge} length and its counterpart in $\mathcal{AN}_2$ to the $graph\_distance$
  \ENDFOR
  \RETURN $ \mathcal{D}(\mathcal{A}_1, \mathcal{A}_2) = graph\_distance$
 \end{algorithmic} 
 \end{algorithm}

\begin{algorithm}
 \caption{\textbf{Computing Graph Distortion Statistics.\\}  We are computing the average distortion, and the standard deviation of distortions, under permutation of nodes. The input is two cliques of the same cardinality, with a mapping from one set of nodes to another.}
 \label{algo:graphstat}
 \begin{algorithmic}[1]
 \renewcommand{\algorithmicrequire}{\textbf{Input:}}
 \renewcommand{\algorithmicensure}{\textbf{Output:}}
 
 \REQUIRE Normalized Adjacency Matrices $\mathcal{N}_1$, $\mathcal{N}_2$  
 of equal dimensions \\
 
 \ENSURE  Baseline for the graph distance, standard deviation of graph distances under permutations of computed distances.

  \STATE Set the value of $graph\_distances$ to an empty list.\\
\vspace*{+2mm}

  \textit{We are permuting the labels of graph, leaving the lengths of the edges intact.}
\vspace*{-5mm}
  \FOR {each permutation $\mathcal{N}_2p$ of the nodes of $\mathcal{N}_2$ }
  \STATE Compute $d = \mathcal{D}(\mathcal{N}_1, \mathcal{N}_2p) $ using Algorithm \ref{algo:distort}
  \STATE Append $d$ to $graph\_distances$
  \ENDFOR
  
  \STATE Set $$graph\_distance\_baseline = Mean( {graph\_distances})$$ 
  $$std = StandardDeviation( {graph\_distances})$$ 
 \RETURN  $graph\_distance\_baseline$, $std$
 \end{algorithmic} 
 \end{algorithm}
 \FloatBarrier

\subsection{Additional experiments}\label{sec:addExperiments}

In the previous section we established that the relatively high similarity between conceptual distances in summary guidelines and full guideline documents was not accidental. 

Tables \ref{tab:Err1} and \ref{tab:Err2} are based on the same type of comparison except based on full guidelines minus \texttt{Unrelated} sentences as described in Section \ref{sec:distauto}. Again we observe that the similarity is not accidental, and that BioASQ embeddings with WM distance seem on average give the best performance.

\begin{table}[h]
\resizebox{\textwidth}{!}{%
\begin{tabular}{|c|c|c|c|c|c|}
\hline
\textbf{Comparing} & \textbf{Min mutual concepts} & \textbf{Model} & \textbf{Distortion} & \textbf{Distortion of permutations} & \textbf{STD} \\ \cline{1-6}
\multirow{18}{*}{Sentence} &\multirow{6}{*}{1} & BioASQ\_CosineSim & 0.526380991 & 0.602890558 & 0.011735664\\ \cline{3-6}
& & Conceptualized\_CosineSim & 0.635564038 & 0.646721788 & 0.011417208\\ \cline{3-6}
& & NoConcept\_CosineSim & 0.626087519 & 0.646906954 & 0.011131221\\ \cline{3-6}
 & & NoConcept\_WMD & 0.352402031 & 0.383852647 & 0.006550777\\ \cline{3-6}
& &  Conceptualized\_WMD & 0.359296888 & 0.390059373 & 0.006626223\\ \cline{3-6}
& &  BioASQ\_WMD & 0.336903254 & 0.384735148 & 0.006498348\\ \cline{2-6}

&\multirow{6}{*}{2} & BioASQ\_CosineSim & 0.449264689 & 0.572620976 & 0.010916054\\ \cline{3-6}
& & Conceptualized\_CosineSim & 0.384945443 & 0.488740293 & 0.008608367\\ \cline{3-6}
& & NoConcept\_CosineSim & 0.433167046 & 0.501788823 & 0.008699466\\ \cline{3-6}
& & NoConcept\_WMD & 0.34284288 & 0.376371094 & 0.006467164\\ \cline{3-6}
& &  Conceptualized\_WMD & 0.330059701 & 0.373155641 & 0.006466969\\ \cline{3-6}
& &  BioASQ\_WMD & 0.32446554 & 0.38365857 & 0.006428759\\ \cline{2-6}
&\multirow{6}{*}{3} & BioASQ\_CosineSim & 0.468163076 & 0.537093759 & 0.010040669\\ \cline{3-6}
& & Conceptualized\_CosineSim & 0.564019791 & 0.57488789 & 0.010091071\\ \cline{3-6}
& & NoConcept\_CosineSim & 0.594326474 & 0.596293202 & 0.010300973\\ \cline{3-6}
& & NoConcept\_WMD & 0.360513492 & 0.375067469 & 0.006461442\\ \cline{3-6}
& &  Conceptualized\_WMD & 0.37193217 & 0.383126986 & 0.006477258\\ \cline{3-6}
& &  BioASQ\_WMD & 0.34276229 & 0.375886963 & 0.006455091\\ \cline{1-6}
\end{tabular}
}%
\caption{
This table shows 
the values obtained in additional experiments, where full document guidelines were modified by attending to concepts in sentences (see Section \ref{sec:dataprep}). 
Column 2, refers to the number of concepts overlapping with summaries. \texttt{Distortion}, shows the distortions of graphs produced using corresponding models from Column 1. As before, in Tab. \ref{tab:featdiff2}, the distortion depends on the how we measure the distances; however,  the shapes of the distributions are very similar.}
\label{tab:Err1}
\end{table}

Note the potentially important observation about Tables \ref{tab:featdiff2}, \ref{tab:Err1} and \ref{tab:Err2}: they jointly show that the property we investigate, i.e. the conceptual distances between guidelines, is indeed geometric, and therefore the word 'distances' is not merely a metaphor. The correspondence between the two graphs is preserved no matter how we set up the experiments. That is, as with geometric properties such as being colinear or parallel, the structure remains the same when a transformation (such as projection) is applied to the points, even though the some of the measurements might change (e.g. measured distances, or area of a parallelogram). The same happens when we transform the documents by removing \texttt{Unrelated} sentences: the values of distortions change, but the non-accidental correspondence with the summary graph (Fig.\ref{fig:graphcompare}) remains invariant.

\begin{table}[]
\resizebox{\textwidth}{!}{%
\begin{tabular}{|c|c|c|c|c|c|}
\hline
\textbf{Comparing} & \textbf{Min mutual concepts} & \textbf{Model} & \textbf{Distortion} & \textbf{Distortion of permutations} & \textbf{std} \\ \cline{1-6}
\multirow{18}{*}{Document} & \multirow{6}{*}{1}
& BioASQ\_WMD & 0.320392721 & 0.38253681 & 0.006475659\\ \cline{3-6}
& & Conceptualized\_WMD & 0.346202932 & 0.389016657 & 0.006561466\\ \cline{3-6}
& & NoConcept\_WMD & 0.351230589 & 0.388633467 & 0.006465622\\ \cline{3-6}

& & BioASQ\_CosineSim & 0.550516174 & 0.534742406 & 0.007178113\\ \cline{3-6}
& & Conceptualized\_CosineSim & 0.568149311 & 0.547872282 & 0.007613218\\ \cline{3-6}
& & NoConcept\_CosineSim & 0.559332088 & 0.54286484 & 0.007445151\\ \cline{2-6}

&\multirow{6}{*}{2} & BioASQ\_WMD & 0.323598367 & 0.386020859 & 0.006486291\\ \cline{3-6}
& & Conceptualized\_WMD & 0.328265638 & 0.378358521 & 0.006481775\\ \cline{3-6}
& & NoConcept\_WMD & 0.341053017 & 0.380095604 & 0.006485268\\ \cline{3-6}
& & BioASQ\_CosineSim & 0.553091569 & 0.536791251 & 0.00725238\\ \cline{3-6}
& & Conceptualized\_CosineSim & 0.558005588 & 0.543056307 & 0.00740679\\ \cline{3-6}
& & NoConcept\_CosineSim & 0.550200164 & 0.539443354 & 0.007298594\\ \cline{2-6}

&\multirow{6}{*}{3} & BioASQ\_WMD & 0.337147756 & 0.38514511 & 0.006439097\\ \cline{3-6}
& & Conceptualized\_WMD & 0.352791102 & 0.37921564 & 0.006506027\\ \cline{3-6}
& & NoConcept\_WMD & 0.351294868 & 0.377266094 & 0.006478541\\ \cline{3-6}

& & BioASQ\_CosineSim & 0.548898679 & 0.536773761 & 0.007261816\\ \cline{3-6}
& & Conceptualized\_CosineSim & 0.555658633 & 0.544321589 & 0.007471369\\ \cline{3-6}
& & NoConcept\_CosineSim & 0.548497913 & 0.540891385 & 0.007362149\\ \cline{1-6}
\end{tabular}
}%
\caption{This table shows the 
values obtained in additional experiments, where full document guidelines were modified by attending to concepts in documents (see Section \ref{sec:dataprep})). 
Column 2, refers to the number of concepts overlapping with summaries. \texttt{Distortion}, shows the distortions of graphs produced using corresponding models from Column 1. As before, in Tab. \ref{tab:featdiff2} and \ref{tab:Err1} the distortion is somewhat depended on the how we measure distances; however, the shapes of the distributions are very similar.}
\label{tab:Err2}
\end{table}

\FloatBarrier
\section{Discussion, future work and conclusions}\label{sec:concls}

Notice that unlike our earlier work \cite{ZadroznyIWCS2017}, in this article we have not performed any logical analysis of the guidelines. Thus the similarities and differences in distances are conceptual, they reflect the conceptual knowledge of the writers (through the word and concept embeddings of their texts), and not the style or formalization of recommendations. 
\textit{{These non-accidental and substantial similarities support our thesis that automated methods can be used for conceptual analysis of guidelines, and in particular can capture some the near-peer epistemic relations discussed earlier. }}\\

We will start our discussion of the results by listing the assumptions that underlie our work; we then argue that the results presented here support the near-peer epistemic model motivating large parts of this research. We conclude with a discussion of some of the limitations of this work and its possible extensions.

\subsection{Our assumptions about modeling epistemic stances using NLP tools}\label{sec:assum}

Before we summarize what we did, and discuss gaps in this work (to be followed up by further research), let's review our assumptions, and put the work in a larger context. 

{This article provides support for the hypothesis that epistemic stances among medical societies can be to a substantial degree analyzed using natural language processing and machine learning tools.} This hypothesis can be decomposed into three ideas, which we discuss briefly below:  
 
\begin{enumerate}

\item Knowledge (i.e. knowledge claims) brought by the authors of guidelines to the writing process is reflected in the concepts they use. 

--- This is intuitively plausible. 

\item This knowledge to a large degree determines the types of recommendations that will be produced. 

--- Again, we can intuitively hypothesize such relation. However, in principle, if the different groups of experts had been epistemic peers
(\cite{lackey2014rel}), i.e. possessed the same knowledge of fact and methods, then, as epistemic peers, they would have produced similar recommendations (if the recommendations were to be inferred from their knowledge, only). As analyzed by the CDC, and shown here in Fig. \ref{fig:cdc}, the recommendations differ. Therefore we either have to reject the complete epistemic peerhood, or assume other influences. The results presented in the sections above strongly suggest some type of near-peerhood. 

\item We assume that vector representations are appropriate for the texts of the guidelines. \\

\end{enumerate}

A few things might be worth keeping in mind, about the above points. Especially, before we discuss near-peerhood and limitations of this work: 

---We prove the near-peer relationship by observing the   geometric similarity between the graph of recommendations and the distances between the vectors representing the full guideline documents (Fig.\ref{fig:graphcompare}).

--- As discussed earlier, in Section \ref{sec:lufig}, given the partial overlaps of the recommendations, the near-peer hypothesis makes sense. In the subsequent sections we have shown these partial overlaps can be recovered from the concepts used in each guideline document. 

--- Note, our vector models cannot account for other factors, e.g. values, different moral utilities, risk perception (\cite{lie2017comparative}), or economic incentives.

\subsection{Modeling epistemic near-peers}\label{sec:conclEpist}

A theoretical model of expert disagreement was proposed in 
\cite{christensenepistemology} and \cite{lackey2014rel}, and analyzed in \cite{garbayo2014} and \cite{Garbayo2018}. Earlier, the epistemology of agreement/disagreement and expertise was discussed in \cite{goldman2001experts}, where the "ideal" model of expertise (e.g. unlimited logical competence) is replaced with a study of situations with epistemic constraints. 
More recently, \cite{grim2019modeling} discusses computational models of epistemology, but does not focus on specifically on disagreement.

The complexity of normative standards in considering truth in the context of disagreement is discussed in \cite{grim2017coherence}. In the data science setting, 
medical expert disagreement and an adjudication process, in analyzing time series data, is described in  \cite{schaekermann2019capturing} and  \cite{schaekermann2019understanding}; there, the authors observe that this process does not eliminate the disagreement, although it reduces its magnitude. Interestingly, the differences in experts backgrounds increase degree of disagreement. This looks to us as another argument for a near-peer model.

The  
epistemic model of multi-criteria expert disagreement \cite{Garbayo2018} can be applied to disagreements among medical guidelines, especially using the near-peer paradigm, which suggests that medical experts  may disagree as a result of being not quite the ideal epistemic peers among themselves.

Accepting the assumptions discussed above in Section \ref{sec:assum} as reasonable, we have shown that epistemic near-peerhood can be modeled using computational tools. That is, that conceptual stances and disagreements among near-peers regarding medical guidelines can be to a large degree observed in the distances between the guideline documents, which are represented as vectors in high dimensional spaces.

Thus, the near-peers share comparable (claims to) knowledge, but also hold relevant differences in their baseline assumptions and sub-areas of expertise and resources. These differences were once idealized in the literature via the epistemic peers model (\cite{lackey2014rel}), but now – as argued in this work --  they can be studied computationally using natural language processing and machine learning tools.

The importance of this new approach lies in its contribution to computational approaches to epistemology (\cite{grim2017coherence}),\  which could provide a complementary representation to the standard formal analysis, represented by Bayesian\footnote{\url{https://plato.stanford.edu/entries/epistemology-bayesian/}}
and formal\footnote{\url{https://plato.stanford.edu/entries/formal-epistemology/}} epistemology.

This might be particularly important for healthcare, intelligence operations, disaster preparedness and others, where there are very tangible consequences of accepting 'wrong' epistemic stances. While what is wrong can only be seen in retrospect, disagreements and differences in epistemic stances can be modeled with deeper (formal) and faster (computational) analysis, as a part of decision modeling.

The experiments reported here show the potential of computational tools to provide a different kind of analysis, and their power to make distinctions not possible before, increasing the resolution of our analysis of disagreements -- just like the microscope allowed human eye to see previously unseen details.

While the intuitive case for positing the near-peer hypothesis is clear, this case would be strengthened by additional experiments. The most obvious ones would be trying replicate our findings for other  guidelines and  other types of expert opinions, .e.g. in public policy. 

Even though our experiments have shown strong influence of concepts on recommendations, the correlation is not perfect, and more accurate models should be possible,  for example, using embeddings from larger models such as the universal encoder (\cite{cer2018universal}) or the GPT family (\cite{radford2019language}, \cite{brown2020language}). 
Another option could be in extending the epistemic model by 
connecting the authors of the guidelines to their other publications, and measuring not the distances between the guideline documents, but explicitly between the groups of authors, represented e.g. by a vector average of their relevant publications. We could then see whether such distances between groups can be translated (in the geometric sense of the word, say as an affine transformation) into distances between the guidelines, and produce a more accurate model.

\subsection{Limitations of this work and future directions}\label{sec:limitats} 
As with any approximate NLP tool, the usefulness of the model depends on its accuracy. In our case, using a very elementary set of tools, we have shown we can approximate the differences in recommendation with about 70\% accuracy. We are sure this accuracy can be improved, and we are actively working towards that goal. In particular, we (or someone else) should be able to use an architecture modeled after our previous system,   \cite{ZadroznyIWCS2017}, to add explicit detection of contradictions to the model. Please note such effort would be highly non-trivial, since the search space of the CDC summary (as in Fig.\ref{fig:cdc}) is much smaller than the search space of the guideline texts in Tab.\ref{tab:guideref}; and additionally the syntax of the actual documents is much more complex than the syntax of the tables.

An obvious extension of this work would be to compare groups of guidelines, e.g.  European medical societies vs. US medical societies. We know that for years their recommendations, e.g. on management of blood cholesterol, differed.

We used word and concept embeddings as a representation of such conceptual stances, but we have not experimented with other representations such as more complex word and document embeddings (\cite{devlin2018bert, peters2018deep, cer2018universal}). Neither have we tried to create more subtle semantic representations based on entity and relationship extraction (e.g. check \cite{NERbioinformatics2017}),  and on formal modeling of contradictions, like the ones discussed in \cite{ZadroznyIWCS2017,zadrozny2018sheaf, garbayo2019dependence}. Such extensions require new methods that would operate, e.g. perform inference, on the document level and not only on the sentence level. Both of these are our active areas of research. 

Another potential limitation of our work has to do with our using only one expert for judgment of conceptual differences between different documents, and leveraging the work of the CDC to tabulate the differences in the recommendations. As discussed in Section \ref{sec:lufig}, given the simplicity of the CDC table in Fig. \ref{fig:cdc}, and with only binary and obvious differences, having only one expert does not seem to be a problem. However,  clearly tabulated comparisons of guidelines, like the ones we started with \cite{CDC_2017} are not common. Thus, translating this work to other set of guidelines will not be trivial. Also, because when working with full text, longer documents, there is more potential for disagreement about building summaries of recommendations. 

There are other  ways of extending the current work. 
In this article we have dealt with a very simple model of near-peer disagreement:  we computed semantic distances between
several pairs of breast cancer prevention guidelines, using different automated
methods, and compared the results an expert opinion.

However, we have not incorporated at any of the logical properties of disagreement in the distance measures, for example unlike our previous work \cite{zadrozny2018sheaf}
\cite{ZadroznyIWCS2017} \cite{garbayo2019dependence} we have not made a distinction between contradictions and disagreements. Thus, some of our ongoing work is about incorporating logical structures of documents and finding metrics reflecting them. \\

\small
\textbf{Description of individual contributions to this research:}
HH performed the majority of experiments discussed in this article, and substantially contributed to writing. 
The idea to investigate computationally the concept of epistemic near-peers comes from LG, as well as the annotations allowing us to use distances, and not just qualitative measures; she also contributed with graphing and overall analysis and writing. SG performed several experiments allowing us to progress with the evaluation. WZ provided the overall supervision, suggested the clique based method for computing distortions, and did the majority of the writing. The overall progress was achieved during multiple team discussions, with equal contributions.

\bibliography{allRefCleaned2020}
\end{document}